\documentclass[manuscript]{acmart}

\AtBeginDocument{%
  \providecommand\BibTeX{{%
    \normalfont B\kern-0.5em{\scshape i\kern-0.25em b}\kern-0.8em\TeX}}}

\copyrightyear{2020}
\acmYear{2020}
\setcopyright{acmcopyright}\acmConference[FDG '20]{International Conference on the Foundations of Digital Games}{September 15--18, 2020}{Bugibba, Malta}
\acmBooktitle{International Conference on the Foundations of Digital Games (FDG '20), September 15--18, 2020, Bugibba, Malta}
\acmPrice{15.00}
\acmDOI{10.1145/3402942.3403011}
\acmISBN{978-1-4503-8807-8/20/09}

\begin{document}


\title{Reflection in Game-Based Learning: A Survey of Programming Games}

\author{Jennifer Villareale}
\affiliation{\institution{Drexel University}}
\email{jmv85@drexel.edu}

\author{Colan F. Biemer}
\affiliation{\institution{Northeastern University}}
\email{biemer.c@husky.neu.edu}

\author{Magy Seif El-Nasr}
\affiliation{\institution{Northeastern University}}
\email{m.seifel-nasr@northeastern.edu}

\author{Jichen Zhu}
\affiliation{\institution{Drexel University}}
\email{jichen@drexel.edu}

\begin{abstract}
Reflection is a critical aspect of the learning process. However, educational games tend to focus on supporting learning concepts rather than supporting reflection. While reflection occurs in educational games, the educational game design and research community can benefit from more knowledge of how to facilitate player reflection through game design. In this paper, we examine educational programming games and analyze how reflection is currently supported. We find that current approaches prioritize accuracy over the individual learning process and often only support reflection post-gameplay. 
Our analysis identifies common reflective features, and we develop a set of open areas for future work. We discuss these promising directions towards engaging the community in developing more mechanics for reflection in educational games.  


\end{abstract}
\maketitle



\section{Introduction}

Reflection, a cycle of "thinking and doing" \cite{Schon1983}, is a crucial component of learning~\cite{boud1996promoting,moon2013reflection,gee2003video, bransford1993ideal}. When learners reflect, the otherwise implicit knowledge becomes digested through active interpretation, questioning, and exploration \cite{fleck2010reflecting,lin1999designing}. As games become an accepted media for education and training \cite{gee2003video, prensky2003digital,randel1992effectiveness,valls2015exploring,honey2011learning}, how well educational games facilitate reflection can have a significant impact on players' learning outcomes. 

There is a significant body of work on game-based learning~\cite{gee2003video,prensky2003digital,randel1992effectiveness}, including areas such as educational content generation~\cite{mott1999towards,hooshyar2018data,valls2017graph}, player engagement and motivation~\cite{sabourin2013affect,wouters2013meta}, learning outcome~\cite{giannakos2013enjoy,spires2011problem}, and personalized learning~\cite{peirce2008adaptive,rowe2009crystal,valls2015exploring}. However, only limited work has been dedicated to reflection in educational games~\cite{moreno2005role,o2014adding}. While reflection occurs in educational games, the educational game design and research community can benefit from more principled knowledge of how to facilitate player reflection through game design. For example, a common approach in educational games is to leave guided reflection to post-gameplay assessments or debriefs \cite{garris2002games,yusoff2009conceptual}, which usually occur separately from gameplay \cite{yusoff2009conceptual}. However, learning science researchers \cite{fleck2010reflecting,lin1999designing} argue that elaboration and explanation will encourage reflection. Instead of merely letting learners revisit the content, it is more helpful to facilitate their reflection on important issues.


Further, these post-gameplay assessments often provide learners with scores or evaluations to measure the correctness of their gameplay \cite{garris2002games,yusoff2009conceptual}. While this may encourage reflection, the focus remains on accuracy over the learning process. Reflecting on process is essential to increase learning outcomes and the learner's awareness of their own learning \cite{lin1999designing,schunk2012learning,edwards2004using}. 
For example, in computer science education, novice programmers must focus less on the correctness of their solution and more on developing their reflective skills on their process. In this way, they can improve their ability to solve new and unseen problems.
Providing players post-gameplay assessments---especially ones that are focused on correctness---may not be enough to motivate reflective behavior at all. Or, if it does, it emphasizes the accuracy of a solution over the learning process \cite{edwards2004using,fekete2000supporting}.

Additionally, due to the diverse work regarding reflection,  it is common that researchers and designers work with different definitions that summarize the act of reflection \cite{fleck2006supporting}. While these definitions may vary across domains and produce different outcomes, they do not address how reflection is facilitated for a variety of learners or how designers can encourage learners toward this ideal state of thinking. As a result, current design approaches incorporate catch-all features that address reflection as a whole \cite{yusoff2009conceptual,garris2002games,hughes2016enhancing}. Often, missing the individual differences that can occur regarding reflection. 


More work is needed to understand what reflection means in educational games, how reflection can be facilitated, and what tools currently exist in games that stimulate reflection. 
Therefore, this paper provides a first step toward identifying common reflective design features used in educational games to develop an understanding of how to better facilitate player reflection through game design. 

In this paper, we explore how reflection has currently been supported in educational games designed to teach programming. We use programming games as our domain of analysis because this genre of educational games is especially well-suited to examine reflection due to the demand for conscious decision making and questioning \cite{lin1999designing}, which are crucial aspects of the reflection process \cite{lin1999designing,fleck2006supporting}. We identified 12 programming games and used them as the basis of our analysis. Our analytical framework contains our prior work on the design space of programming games\cite{zhu2019programming} and two theoretical frameworks on reflection including 1) Sch\"{o}n’s notions of reflection-in-action and reflection-on-action, to specify when reflection features occur, and 2)  Lin et al.'s four reflective design features \cite{lin1999designing} to specify what reflective elements occur in each game. 

This paper presents our analysis of the 12 programming games. Using a close reading approach \cite{bizzocchi2011well}, we identified common reflective design features used in the games. We found that these games provided reflective features; however, their features were often prioritizing the accuracy of a solution after gameplay. We also identify promising directions towards developing a wider range of design features specifically geared towards supporting reflection in learning games. 

The rest of the paper is organized as follows. We begin by providing an overview of the literature and existing work on reflection. Then, we follow this by elaborating on our methodology to identify reflective features. We then detail our findings for each feature and conclude the paper with possible directions towards more mechanics for reflection.

\section{Related Work}
\label{rw}
In this section, we provide an overview of the literature on reflection and focus on the learning science perspective. 
Then, we discuss the importance of reflection in Computer Science, specifically, reflective practices when learning how to program due to our focus on programming games. Finally, we pay special attention to existing work on what elements in technology design, including in educational game design, can facilitate and guide reflection. 

\subsection{Reflection and Its Impact on Learning}
Reflection has been an active topic in learning science research \cite{schunk2012learning, Schon1983, gee2003video, boud1996promoting, moon2013reflection, bransford1993ideal}. Researchers have defined reflection as a process where one can engage in an exploration of experiences that lead to new understandings and appreciations \cite{boud1996promoting}. 
It has also been described as an essential tool to allow learners to make conscious value choices in their practices \cite{boud1996promoting} and enable them to adapt their thinking to other situations \cite{lin1999designing,fleck2010reflecting}. 
It is widely accepted that reflection is not only beneficial but also necessary for learning \cite{schunk2012learning, Schon1983, gee2003video, boud1996promoting, moon2013reflection, lin1999designing}. In reflective practice, a learner becomes aware of otherwise implicit knowledge or behavior \cite{Schon1983,sengers2005reflective}. Helping learners to reflect can lead to a variety of benefits, such as self-development, decisions or resolutions of uncertainty; empowerment or emancipation; and other outcomes that are unexpected \cite{moon2013reflection}. Once learners become efficient at reflection, they have more potential to become effective lifelong learners and adapt to new situations \cite{lin1999designing, sengers2005reflective}.

Donald Sch\"{o}n's \cite{Schon1983} work in the \textit{Reflective Practitioner} is one of the most widely cited works regarding reflection. He describes two types of reflection, in-action and on-action. Reflection-in-action is the process of revisiting practices during action through "thinking and doing" or testing out a hypothesis in a new environment and using the results to drive the next action. For example, one can be reflective during the action as it is happening, such as playing a song with an instrument and reflecting on how the sound can change. Reflection-on-action is the process of reviewing or analyzing a situation after an event. For example, listening to a recording of the finished song, after the action is done and reviewing it in comparison to other performances. Both of these processes feed into each other, creating a cycle of reflection during and after an event \cite{Schon1983}. Reflection-in-action extends thinking into reflection-on-action, and the results motivate the following actions. Therefore, reflection should not be thought of as a separate activity from action, but as an essential part of the overall experience \cite{sengers2005reflective, Schon1983}. This framework is particularly useful because it helps identify when reflective elements occur in our game analysis.

\subsection{Reflection and its Impact on Programming}
Reflection is a central skill when learning how to program and has been an active topic in computer science education \cite{edwards2004using,fekete2000supporting}. Reflective practices are especially important in an environment when there exists more than one approach to solve a problem. For example, CS students must be reflective when finding a viable solution, especially when previous programs developed may not work in a new context. Therefore, analytical skills are needed to better equip CS students on how to go about learning successfully \cite{fekete2000supporting}. 

Edwards \cite{edwards2004using} explicitly calls for the need to encourage CS students to reflect-in-action when developing code. Novice programmers usually use a "trial and error" strategy when developing programs to see what works and what does not. This behavior is further reinforced by the way current programming assignments in CS education are structured. Assignments tend to emphasize output correctness when students receive feedback only on the end result they produce. Therefore, students equate a program that "produces the right output" with an "effective solution" \cite{edwards2004using}. Edwards suggests that students must shift to more reflective behaviors where assignments encourage hypothesis-forming and experimental validation to develop students' analytical skills. 
This issue can also be observed in existing programming games where reflective features, such as post-gameplay assessments, focus on outcomes and the correctness of a solution at the end of each level. To help encourage students to be more reflective in their learning process, Edwards proposes a test-driven development approach that evaluates student performance on how well they have demonstrated the correctness of their program through testing. The goal is to support rapid cycling between writing individual tests and adding small pieces of code, incrementally "thinking and doing." This approach has the potential to increase reflection-in-action in programming games by developing more frequent performance feedback throughout the game experience toward the learning process.

\subsection{Designing for Player Reflection}
Despite its crucial impact on learning, it is generally difficult for learners to be reflective on their own \cite{fleck2010reflecting}.
A common approach in educational games is to display assessments of students learning outcomes. For example, using debriefs to describe important details of the experience, such as a description of events that occurred, a discussion of mistakes, and corrective actions \cite{garris2002games,yusoff2009conceptual}. 
However, merely revisiting content without elaboration or a reason is not reflection \cite{fleck2010reflecting}.
Some work addresses this issue by asking students to provide explanations for their answers \cite{moreno2005role,o2014adding}; however, if the answer is incorrect, this may deepen existing misconceptions and may not facilitate reflection toward essential issues at all.
It is important for reflectors to be encouraged to think specifically about issues that are considered important \cite{fleck2010reflecting}. 
As a result, providing assessments and space for explanation may not be enough to motivate valuable reflective behavior.

To the best of our knowledge, there have been no existing attempts to analyze reflection in educational programming games. Therefore, we widened our scope to look into digital environments that include reflection as part of the design.
Previous work addresses reflection in games, such as frameworks that organize reflection into the gameplay cycle as a post-gameplay assessments or debriefs \cite{garris2002games,yusoff2009conceptual}, or facilitates reflection through self-explanations, questionnaires \cite{o2014adding,moreno2005role}, and in-game journals \cite{maciuszek2011computer}.
In the context of educational games, several pieces of work discuss how to design for reflection. For instance, Yusoff et al. \cite{yusoff2009conceptual} proposed a conceptual framework for serious games on the basis of nine different principles, with reflection being one \cite{yusoff2009conceptual}. They based this reflection principle on previous work from Garris et al. \cite{garris2002games} which discusses using a debrief, a review and analysis of events that occurred in the game, to provide a link between the game cycle and learning outcomes. 
Yusoff et al. \cite{yusoff2009conceptual} highlighted this work in their framework by creating a separate reflection activity that provides players a description, an explanation of why this activity is chosen, a discussion of the errors made by the learner, and corrective suggestions after gameplay.  

Existing work in educational games have focused on using written responses to prompt reflection through explanations.
Moreno et al. \cite{moreno2005role} investigated guidance and reflection in an interactive multimedia game, Design-A-Plant. 
The goal of their set of studies is to pinpoint the role of guidance and reflection in promoting scientific understanding in agent-based multimedia games for college students. Regarding reflection, in-game agents would prompt an elaborative interrogation, where the agent would ask the student to provide an explanation for their answer during a problem-solving session. 
Initial results showed that asking students to reflect through explanations on their answers did not affect learning. Moreno et al. \cite{moreno2005role} suggest that the student's answers could be incorrect and may promote the consolidation of an incorrect mental model by having students verbalize their misconceptions. In a later study, students were asked to reflect on a correct solution rather than their own, which then showed positive results regarding retention and transfer measures.

O'Neil et al. \cite{o2014adding} studied the impact of self-explanation prompts in a math game. These prompts are requests for players to type an explanation or to select a reason from a list for each move they made to encourage reflection. This study showed that explanation prompts could help learners make connections between game terminology and mathematics terminology. They found explanation prompts to be an effective instructional method. In some cases, the prompts were less useful when very simple or very abstract questions were asked.
Also, the authors do express concern that self-explanation prompts "slow learners down" during the game. This agrees with work from O'Neil et al. which states that prompts were most effective when they were designed to not distract from learning \cite{o2014adding}.

Additional reflection work can be found in educational tools for teachers.
Hughes et al. \cite{hughes2016enhancing} incorporates a reflective activity called, after-action review (AAR), in a simulated learning environment for teachers. 
This is a similar approach we see in existing educational games to encourage reflection. 
AAR provides users time to reflect and guidance with video tagging software and a playback \cite{dieker2017using,hughes2016enhancing}. Additionally, audio and gesture capture accompanies a post-experience assessment of how the user handled different situations during the session \cite{barmaki2018embodiment}.
As noted previously, displaying assessments post-experience may not be enough to have learners be reflective on their own, especially toward their learning process. More work is needed on how to encourage and guide learners through the cyclic process of reflection, both in-action and on-action.

Technology design has also explored how we can develop interactive environments to encourage reflective behaviors.
The framework we find particularly useful to further the discussion on how to design for reflection in educational games is from Lin et al. \cite{lin1999designing}. Developed for learning technology in general, this theory discusses how technology can be designed to support reflection, the same objective as this paper, but we narrow our focus to educational games. They describe how reflection can be incorporated in a variety of learning environments and the conditions to facilitate reflection. 
As a result, this framework is easily transferred to the context of educational games because they provide specific examples that are found in existing technology design. The features include: (1) process displays, (2) process prompts, (3) process models, and (4) social discourse. Each feature will be discussed further below. 
In addition, Lin et al. \cite{lin1999designing} state that technology provides a new set of opportunities for reflection, "technology, properly designed and used, enables us to realize reflective learning environments that were not previously possible. (pg.44)" By that logic, games are an applicable domain for new reflective thinking and design. 

Another useful framework to further the discussion is Fleck et al.'s \cite{fleck2010reflecting} Levels of Reflection. These levels incorporate reflection into a behavioral framework by examining purposes of reflection, conditions for reflection, and details of how technologies can support these different levels. 
The four levels of reflection are R1) Reflective Description, revisiting material; R2) Dialogic Reflection, exploring relationships; R3) Transformative Reflection, revisiting with the intent to do something different; and R4) Critical Reflection, considering aspects beyond the immediate context. We do not use this framework in our analysis because behaviors and activities associated with reflection was the focus. Our work, instead, focuses on design features in programming games that stimulate reflection. However, Fleck et al. make an important argument that we wish to emphasize that we should build technologies to support multiple levels of reflection. To do this, designers should consider various behaviors of their users, which Fleck conveniently organized into levels. Unfortunately, we still do not know how to encourage reflection toward these levels, particularly in games.

\section{Mechanics for Reflection in Games}
To advance the knowledge of how to design for reflection, we conducted an exploration of reflection in programming games and what features currently exist that stimulate reflection. 
We narrow our focus to a sub-genre of this domain, programming games, to start with a smaller sample for our analysis.
Additionally, programming games are one of the sub-genres of this domain that are especially well-suited to examine reflection because they tend to embed complex problems that demand conscious decision making and questioning \cite{lin1999designing}. Following our analysis, we discuss common reflective patterns to investigate open areas for future work in reflection design.

\subsection{Methodology} \label{meth}

In this paper, we utilize Lin et al.'s \cite{lin1999designing} four Reflective Design Features to identify which reflective elements occur and Donald Sch\"{o}n’s \cite{Schon1983} in-action and on-action framework to specify when these features occur, during gameplay or post-gameplay.
We define each feature and provide examples in section \ref{results}.
We chose to use these two frameworks due to their focus on design features, which aided in identifying what reflection features occur in technology design and learning environments. 
Lin et al.'s framework offered specific features found in existing technology that was applicable to transfer to game environments for our analysis.

We select these games based on existing work in educational programming games \cite{zhu2019programming} 
and on popularity identified with web searches. This led to the selection of the following games: {\em Blockly: Maze} \cite{blockmaze}, {\em Cargo-Bot} \cite{cargobot}, {\em Code Combat} \cite{codecombat}, {\em CodeMonkey} \cite{codemonkey}, {\em Code Spells} \cite{codespells}, {\em CodinGame} \cite{codingame}, {\em Human Resource Machine} \cite{humanresource}, {\em Light Bot} \cite{lightbot}, {\em Manufactoria} \cite{manufactoria}, {\em Parallel} \cite{zhu2019programming}, {\em RoboZZle} \cite{roboZZle} and {\em SpaceChem} \cite{spacechem}.


Two researchers developed a shared document that included the feature, its definition, and examples of how this feature was developed in existing technology. All examples and descriptions that were used are listed in the following sections to ensure consistency. Both researchers collectively played each game via a close reading approach \cite{bizzocchi2011well}, where they took notes—during and after play—about the specifics of what and when reflection features occur. Frequently, comparing their notes and assessments to the feature document. If a feature was unclear, both researchers would interact with the game in question and agree after a discussion. Each game's notes were shared and discussed by both researchers to conclude the results.

\begin{figure*}[!t]
  \includegraphics[width=\textwidth]{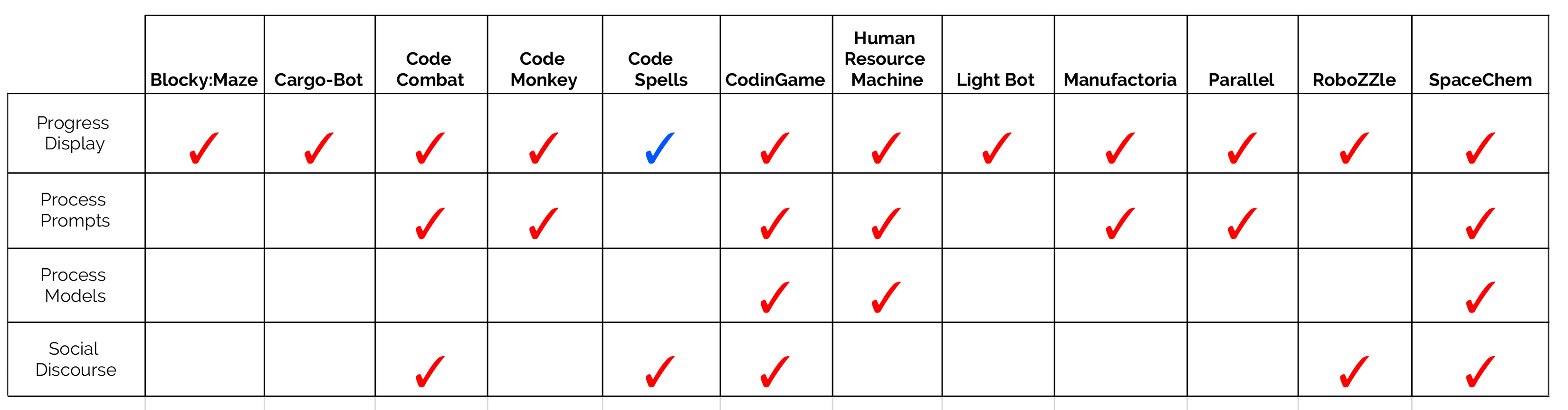}
  \caption{Reflective Design Patterns in 12 Analyzed Programming Games. A check mark indicates that the reflection feature occurred in the game. Blue: reflection-in-action; Red: reflection-on-action.}
\label{fig:patterns}
\end{figure*}

\section{Results} \label{results}
Based on our analysis of 12 programming games, we identify what reflective features are being used and when they occur in each game. Figure \ref{fig:patterns} shows which games had a particular reflection feature and when they occurred. Next, we define, provide examples, and detail our findings for each feature.



\subsection{Process Displays}
\textit{Process displays} are displays that visualize problem-solving and thinking processes. This feature traces or records students' processes or actions inside the environment. Then, these actions are visualized in the interface that reflects the process back to the students to examine.
For example, Geometry Tutor was designed to teach students geometry. The program used process displays to trace, record, and visualize students' actions. The program visually shows the students' action paths in inspectable tree diagrams to allow students to examine their process in more detail  \cite{lin1999designing}.


Based on our analysis, 11 out of 12 games provided an in-process playback to visualize the results of their solutions on-action. Often these displays occur on a local, level-by-level basis.
All of these games provided a way to visualize the player's program, solution, or creation by playing back their solution in the context of the game's environment and objective. However, these playbacks focuses on output correctness and did not explicitly encourage thinking toward the individual player's learning process.
Only one game, \textit{Code Spells} \cite{codespells}, did not use an in-progress visualization or focus on any evaluative outcome; rather, they demonstrated the player's solution in-action in an open world environment.
Further, some of these games had a programming space and a game space. The programming space is where the player codes or places components. The game space is where the code can be visualized in the context of the game \cite{zhu2019programming}. We use these spaces to describe the variations of process display.

We observed two ways for process display to occur, either in a split view or in a combined view of the programming and game space. 
Of the 11 games only 3, \textit{SpaceChem} \cite{spacechem}, \textit{Manufactoria} \cite{manufactoria}, and \textit{Parallel} \cite{zhu2019programming}, blended both spaces so the player would program directly onto the game space. 
Alternatively, the other games all used a split view and had separate spaces, one for programming and another for observing the playback. 
As a result, they handled the process display feature differently, but still provided a playback for players to reflect-on-action and observe the results of their creations.

\begin{figure*}[!t]
  \includegraphics[width=\textwidth]{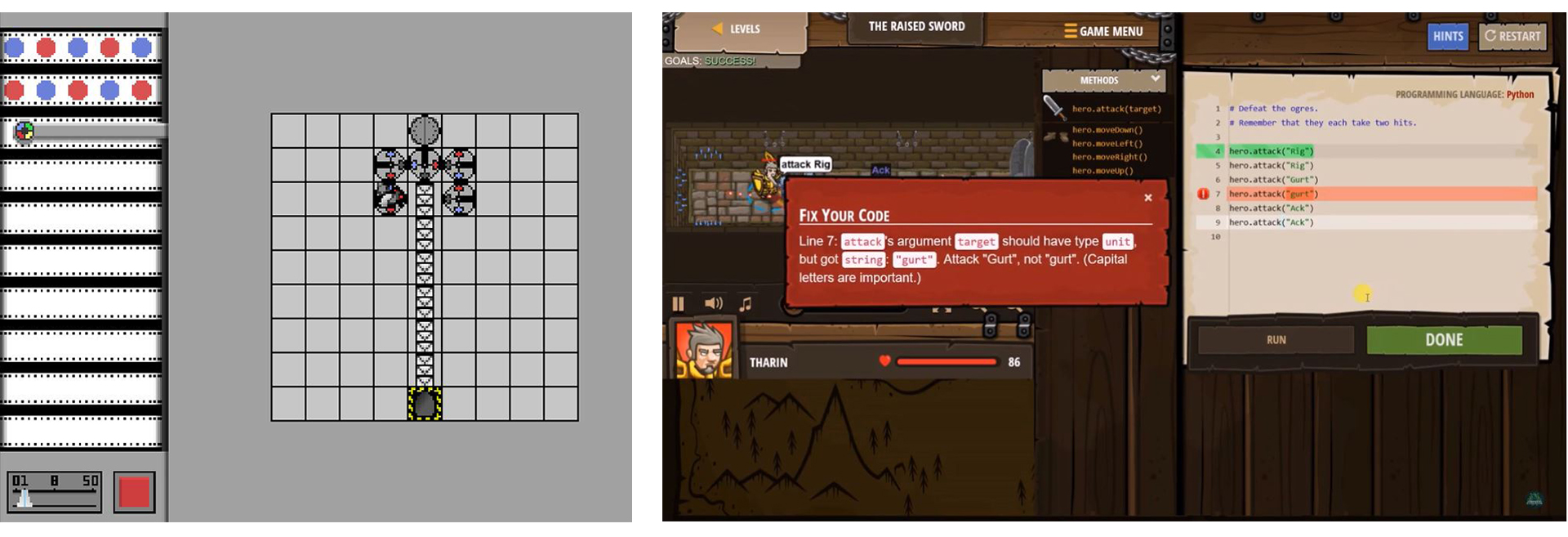}
  \caption{Manufactoria (Left) uses Process Display by visualizing the player's solution in a playback. CodeCombat (right) uses a Process Display to highlight the player's code, line-by-line and visualize the result of a players code in a playback. This game also uses Process Prompts on failure to encourage reflection on incorrect code.}
\label{fig:mancom}
\end{figure*}

For the split view, the player can hit submit to see their solution visualized in the game space. Most of these games use a highlight that steps through the player's code line-by-line. This mechanic is more of a "point" which showcases what code is currently active in the playback. 
For example, \textit{CodeCombat} \cite{codecombat}, is a game about software programming concepts and languages. In this game, players write code to command a hero through a variety of battles. In each level, the players can hit run to play their solution and see their hero enact the written code.
As seen in figure \ref{fig:mancom}, the game highlights the code line-by-line to showcase what code is being used by the hero. This visualization encourages reflection-on-action because it steps through the player's process and provides an opportunity to observe the results of each line of code. While this plays, the player can consider their process while observing the results. However, as we will discuss further in the next section, on an error, the playback stops and provides a process prompt to indicate where the player has made an error. 

For the combined view, the player could still visualize their solution in the game space; however, there is no highlighting mechanic. After players finish their solution, the game space would show their solution in action.
For example, \textit{Manufactoria} \cite{manufactoria}, 
as seen in figure \ref{fig:mancom}, the player can drag components onto the track to build directly in the game space. \textit{Manufactoria} \cite{manufactoria} provides a play button for players to observe the outcome of their machines in the level's scenario. 
This encourages reflection-on-action by providing an opportunity to observe their finished machine. Since this view is combined, players can easily cycle between building and observing since they do not have to switch between spaces.

\begin{figure*}[!t]
  \includegraphics[width=\textwidth]{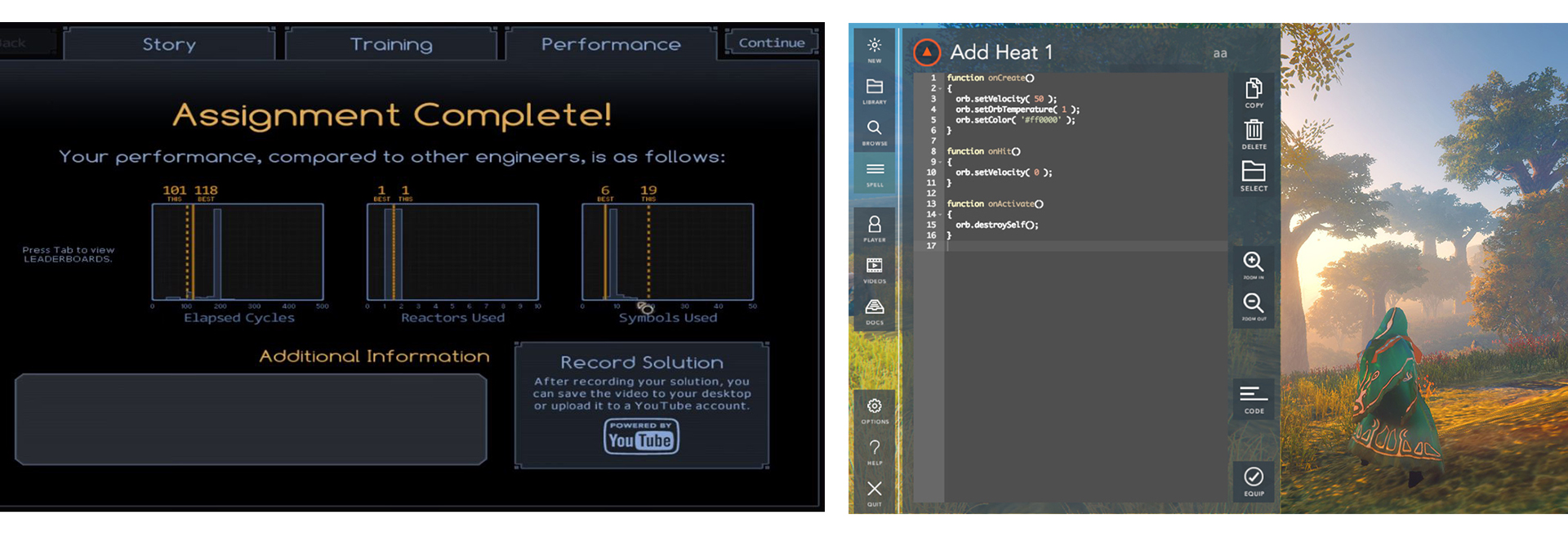}
  \caption{SpaceChem (Left) uses Process Models by comparing the player's score to other players. Code Spells (Right) uses Social Discourse by incorporating a community library where players can share  code.}
\label{fig:spacespell}
\end{figure*}

Only one game, \textit{Code Spells} \cite{codespells}, did not use an in-progress visualization. Instead, they demonstrated the player's solution in-action where the player could use their creations in an open world environment.
\textit{Code Spells} is an adventure game where players use computer code and programming to act within the game world.
Players can create, modify, or edit code in the programming space. They can cast their created spells and navigate the world in the game space, as seen in \ref{fig:spacespell}. Both spaces, coding and exploring, potentially encourages more reflection-in-action because the player is active in both areas. 
This game uniquely incorporates the ability to test out the player's code in the game actively. They can play with and test their spell in a variety of different environments as they explore the game world, thus there are more opportunities to reflect-in-action toward building hypotheses and testing them in both spaces. 

\subsection{Process Prompts}
\textit{Process prompts} are UI elements, such as pop-ups, windows, or pages in a digital interface that poses appropriate questions or content to help guide students in tracking and understanding their own process. 
Process prompts can be used to encourage reflection by directly asking students questions and requiring them to explain their actions.
For a example, Isopod Simulation Program uses process prompts to ask students questions and to encourage students to explain their thought process through a popup window with text.
They can also be used to encourage reflection indirectly by helping students generate their own questions through assessments of their work against the given criteria.
For an example, CIR-CSIM-Tutor program uses process prompts to aid student reflection on error. This window displays content specific tasks and assesses the student's work against the given criteria \cite{lin1999designing}.

    
From our analysis, the most common time for a process prompt to occur was on failure. 7 out of 12 games used process prompts to aid student reflection on an error by highlighting specific content in the level, assessing the player against the level's objectives, or displaying helpful tips or corrective content. 
Since the focus of these prompts were on the correctness of the solution, these prompts drew attention to the level's evaluative outcomes. No prompts were used to guide reflection on the learning process of building a solution. Rather, it highlighted what needed to be fixed before moving to the next level.

We observed two ways that process prompts were used to facilitate reflection-on-action: 1) to a specific area and 2) to the overall goal of the level.
\textit{CodeCombat} \cite{codecombat} is a game about software programming concepts and languages.  When the player is finished writing their code, the player can hit run to play their solution and see their hero enact the written code line-by-line through a process display feature.
However, if the system encounters an error, a popup message appears along side the incorrect code. As seen in figure \ref{fig:mancom}, it uses process prompts to direct player reflection to a specific area that has an error. This popup displays content with corrective details and includes fixes for the player to consider. This prompt encourages reflection-on-action toward a specific area of importance. Once the code is fixed, the player can continue to the next level.

\textit{Parallel} \cite{valls2017graph,alderfer2018lessons} is a game designed to teach parallel programming. In this game, the player needs to coordinate multiple arrows on a track by picking up packages and delivering them to their destination. This game uses process prompts to encourage player reflection on the level's objectives and goals on a failed solution. When the player submits a solution, the visualization allows the player to observe the incorrect solution play out. If the level goals were not met, the game displays a popup menu with an assessment of their solution against the level's objectives. This encourages reflection-on-action toward the level's objectives.


\subsection{Process Models}
\textit{Process models} are displays that showcase experts' thinking processes so students can compare and contrast with their own process. 
For example, ASK Jasper, a computer-based learning environment, teaches students basic geometry through designing a swing set. Students can compare and contrast their own designs and reflect on the design process of an expert \cite{lin1999designing}.

Process Models display other expert solutions or actions in a learning environment through pages or other windows. Often these occur after the learner has first completed their solution on their own. We found that 3 out of the 12 games use process models to facilitate reflection on another player's solution through a leader board or comparative stat menus. None of the games incorporated an explicit expert to model a particular process in the game. \textit{CodinGame} \cite{codingame} used a community voted leader board which allows players to view higher voted solutions.
This activity takes place outside the game on a separate web page and does not incorporate this model directly into the game.
Even though the game does not display how an expert would solve a level, players can observe other player's code based on what the community voted as the top solution. 
Since we did not observe any expert models in our analysis, we focus on how players compare and contrast their own process with another to locate where these expert models could potentially go into educational games.

Process models display what others have done at the same level or, more often, include a more distilled version of the solution with a comparative stat menu.
For example, \textit{SpaceChem} \cite{spacechem}, see figure \ref{fig:spacespell} is a game that teaches concepts related to both chemistry and programming by using a visual programming language. In this game, players build chemical molecules via an assembly line by programming a variety of components. 
\textit{SpaceChem} \cite{spacechem} uses process models to encourage reflection by using a popup menu with a histogram. It facilitates reflection on the player's stats against the general average. This display allows players to see the average time and number of parts used before they start the level. After a successful solution has been reached, the popup menu appears again to show where the player stands in comparison to others. These do not provide expert answers, but players are provided with a "compare and contrast" reflection opportunity that encourages players to revisit and improve their solution.

\subsection{Social Discourse}

\textit{Social Discourse} is a space in the interface for community-based discussion where students can examine multiple perspectives and receive feedback on their process that can then be used for reflection. This feature refers to the collaborative nature of reflection when a learner seeks input from a group to modify their practice. 
For example, CSILE is a program for students to collaborate. In this program, students can discuss questions and collaborate to understanding a problem \cite{lin1999designing}.

Social discourse allows learners to participate in discussions and provide feedback on open problems. Additionally, this offers different perspectives on the same problem.
We found that 5 out of 12 games use a social discourse feature in a shared content library or in an external website. However, none of the games we analyzed incorporated an area for discussions inside of the game. 
Only one game, \textit{Code Spells} \cite{codespells}, incorporated a social feature in which players could share code. 
As seen in figure \ref{fig:spacespell}, this game uses a form of social discourse to encourage reflection on other spells through a library player submitted code. It facilitates reflection on different approaches to coding and thinking through a solution.
The game's interface provides access to a community library of other published spells. Players can view, edit or reference spells that other players have published. However, there is no direct communication in the game.
Reflection occurs on-action by players evaluating other player's code. Even though this game did not include a space for discussion, it displayed an approach to encourage collaboration through participation in a game library.

Another social feature we observed included conversations, such as how others would approach a problem, and questions in which the community can provide answers and feedback to the player. It encourages reflection on multiple perspectives. However, this occurs outside of the game on a website.
\textit{CodinGame} \cite{codingame} uses social discourse to promote reflection on other player's solutions through web pages separate from the puzzle. 
The website displays voted solutions that form a leader board. They also include a space for conversations under each solution and in a forum in which people can discuss puzzles and provide feedback to other puzzles.

\section{Discussion} \label{toward}
In the previous sections, we present our analysis of 12 programming games and detail what mechanics have been used to stimulate reflection. Based on our observations, all 12 games demonstrated reflective features; however, we observed reoccurring designs that expose a lack of reflection mechanics in educational games. 

We observed only one game, \textit{Code Spells} \cite{codespells}, use a reflective feature during gameplay (reflection-in-action), as most of the games provided post-gameplay assessments (reflection-on-action), separate from the gameplay. This disparity is inconsistent with existing learning science theory that argues reflection needs to take place both during and after an action \cite{Schon1983}. While considering learning science theories, this can be a useful starting point to encourage new forms of reflection mechanics that have yet to be explored in educational games.

While all 12 games used some form of process display, some games only offered playback of learners' actions. Based on Lin et al.'s work \cite{lin1999designing}, it would be useful for these games to add guided prompts, both in random and fixed presentation schedules, so that learners' attention can be drawn on specific aspects of their learning process. For example, in \textit{CodeCombat} \cite{codecombat}, we could add guided prompts to the solution playback as a way to direct attention to specific areas or ask players to elaborate on a solution, as opposed to only prompting the player on error. 


Below we describe our observations regarding the common design patterns found from our analysis and suggest open areas for future work. We discuss these promising directions towards engaging the community in developing more mechanics for reflection.

{\bf {\em Supporting reflection-in-action.}} The first pattern we observed from our analysis was that very few games had reflective features during gameplay and often left reflection to post-gameplay assessments or debriefs. As a result, supporting the whole process reflection, especially during gameplay, is a significantly underexplored area in programming games.  

This design choice may indicate designers not wanting to overwhelm or interrupt the gameplay experience, as some of the more complex puzzle games already require a significant amount of thinking and doing to complete the level. However, by not considering reflection features during gameplay, players may be more inclined to focus on the accuracy of the solution, or only on getting a correct answer, as opposed to building an effective solution. Reflection support is especially needed in programming games as a solution may work in one setting, but may not always work in a new context. Encouraging more frequent feedback during gameplay may help direct reflection on the learning process.

A possible direction is to consider reflection as its own gameplay loop and how it evolves during and after each level, similar to any other in-game skill. For reflection-in-action, incrementally using process display and process prompts during gameplay may support a consistent reflection cycle and direct thinking toward important areas. We suggest allowing the player to edit their solution while the in-progress visualization is occurring. This may provide the player with a more active role in these playbacks. Process prompts can also be used to support frequent feedback and direction toward what the player may need to reflect on specifically.

For reflection-on-action, a possible direction is to incorporate a social discourse feature to highlight different perspectives on a problem. This feature may emphasize process over outcome, as this can showcase a variety of approaches rather than displaying the average scores as this may further reinforce thinking on accuracy. A social feature can also promote the collaborative nature of reflection when a player examines and shares different approaches in the community.

{\bf {\em Reflection from a global perspective.}} The second pattern we observed from our analysis was that all refection features were confined to local, level-by-level content. None of the features we surveyed provided a higher viewpoint for players to reflect on their actions across multiple levels or problems. This pattern is another underexplored area for reflection work.  

By visualizing the player's gameplay over numerous levels, it may provide more opportunities for players to reflect on their behaviors that may otherwise be missed from a local perspective, such as common strategies or approaches. A possible direction is to use a process display feature to showcase gameplay strategies over multiple levels or attempts on a solution. Additionally, providing an opportunity for the player to compare process displays with other players may also emphasize the differences in an approach, further highlighting otherwise unnoticed behaviors.

{\bf {\em Multiple forms of reflection.}} The third pattern we observed was that the reflection features were always consistent for each player. This consistency provides a cohesive player experience but does not address the differences in reflection. For example, a process prompt may encourage reflection in one player, but may not in another. Exploring multiple forms of reflection is another underexplored area in educational games.

Based on work in section \ref{rw}, reflective behaviors can vary depending on the individual. It is crucial to develop this work further if educational games are to ensure every player gains the benefits of reflection to improve their learning outcome. As a research community, we need to better understand how reflection is facilitated for different people and how games can support these differences. 

Existing work by Fleck et al. \cite{fleck2006supporting} has already established different levels of reflection. Yet, we still do not know the different ways to facilitate reflection toward these desired levels. A user study is warranted to explore these individual differences and how game design can be used to support reflection further.

\section{Conclusions and Future Work}
In this paper, we explore reflection in 12 programming games and detail what features have been used to stimulate reflection. We identify existing reflection mechanics in programming games and what tools currently exist to facilitate this behavior. As a result of our analysis, we developed a set of common reflective patterns and suggested open areas for future work. We discuss these promising directions towards engaging the community in developing more mechanics for reflection in educational games. A limitation in our work is our small sample of games, and we encourage researchers to extend this work to a larger sample and investigation into other genres of educational games. Finally, we note the subjectivity of our results and discussion, and we support different interpretations of this topic. 


\begin{acks}
This work is partially supported by the National Science Foundation under Grant Number 1917855. The authors would like to thank all current members of the project.
\end{acks}

\bibliographystyle{ACM-Reference-Format}
\bibliography{references} 


\begin{thebibliography}{47}


\ifx \showCODEN    \undefined \def \showCODEN     #1{\unskip}     \fi
\ifx \showDOI      \undefined \def \showDOI       #1{#1}\fi
\ifx \showISBNx    \undefined \def \showISBNx     #1{\unskip}     \fi
\ifx \showISBNxiii \undefined \def \showISBNxiii  #1{\unskip}     \fi
\ifx \showISSN     \undefined \def \showISSN      #1{\unskip}     \fi
\ifx \showLCCN     \undefined \def \showLCCN      #1{\unskip}     \fi
\ifx \shownote     \undefined \def \shownote      #1{#1}          \fi
\ifx \showarticletitle \undefined \def \showarticletitle #1{#1}   \fi
\ifx \showURL      \undefined \def \showURL       {\relax}        \fi
\providecommand\bibfield[2]{#2}
\providecommand\bibinfo[2]{#2}
\providecommand\natexlab[1]{#1}
\providecommand\showeprint[2][]{arXiv:#2}

\bibitem[\protect\citeauthoryear{Alderfer, Smith, Onta{\~n}{\'o}n, Char,
  Nebolsky, Zhu, Furqan, Freed, Patterson, and Valls-Vargas}{Alderfer
  et~al\mbox{.}}{2018}]%
        {alderfer2018lessons}
\bibfield{author}{\bibinfo{person}{Katelyn~Bright Alderfer},
  \bibinfo{person}{Brian~K Smith}, \bibinfo{person}{Santiago Onta{\~n}{\'o}n},
  \bibinfo{person}{Bruce Char}, \bibinfo{person}{Jessica Nebolsky},
  \bibinfo{person}{Jichen Zhu}, \bibinfo{person}{Anushay Furqan},
  \bibinfo{person}{Evan Freed}, \bibinfo{person}{Justin Patterson}, {and}
  \bibinfo{person}{Josep Valls-Vargas}.} \bibinfo{year}{2018}\natexlab{}.
\newblock \showarticletitle{Lessons Learned From an Interactive Educational
  Computer Game About Concurrent Programming}. In
  \bibinfo{booktitle}{\emph{Proceedings of the 49th ACM Technical Symposium on
  Computer Science Education}}. \bibinfo{publisher}{Proceedings of the 49th ACM
  Technical Symposium on Computer Science Education},
  \bibinfo{pages}{1077--1077}.
\newblock


\bibitem[\protect\citeauthoryear{Barmaki and Hughes}{Barmaki and
  Hughes}{2018}]%
        {barmaki2018embodiment}
\bibfield{author}{\bibinfo{person}{Roghayeh Barmaki} {and}
  \bibinfo{person}{Charles~E Hughes}.} \bibinfo{year}{2018}\natexlab{}.
\newblock \showarticletitle{Embodiment analytics of practicing teachers in a
  virtual immersive environment}.
\newblock \bibinfo{journal}{\emph{Journal of Computer Assisted Learning}}
  \bibinfo{volume}{34}, \bibinfo{number}{4} (\bibinfo{year}{2018}),
  \bibinfo{pages}{387--396}.
\newblock


\bibitem[\protect\citeauthoryear{Bizzocchi and Tanenbaum}{Bizzocchi and
  Tanenbaum}{2011}]%
        {bizzocchi2011well}
\bibfield{author}{\bibinfo{person}{Jim Bizzocchi} {and} \bibinfo{person}{Joshua
  Tanenbaum}.} \bibinfo{year}{2011}\natexlab{}.
\newblock \showarticletitle{Well read: Applying close reading techniques to
  gameplay experiences}.
\newblock \bibinfo{journal}{\emph{Well Played 3.0: Video Games, Value and
  Meaning}} (\bibinfo{year}{2011}), \bibinfo{pages}{262--290}.
\newblock


\bibitem[\protect\citeauthoryear{Boud, Keogh, and Walker}{Boud
  et~al\mbox{.}}{1996}]%
        {boud1996promoting}
\bibfield{author}{\bibinfo{person}{David Boud}, \bibinfo{person}{Rosemary
  Keogh}, {and} \bibinfo{person}{David Walker}.}
  \bibinfo{year}{1996}\natexlab{}.
\newblock \showarticletitle{Promoting reflection in learning: A model}.
\newblock \bibinfo{journal}{\emph{Boundaries of adult learning}}
  \bibinfo{volume}{1} (\bibinfo{year}{1996}), \bibinfo{pages}{32--56}.
\newblock


\bibitem[\protect\citeauthoryear{Bransford and Stein}{Bransford and
  Stein}{1993}]%
        {bransford1993ideal}
\bibfield{author}{\bibinfo{person}{John~D Bransford} {and}
  \bibinfo{person}{Barry~S Stein}.} \bibinfo{year}{1993}\natexlab{}.
\newblock \showarticletitle{The IDEAL problem solver}.
\newblock  (\bibinfo{year}{1993}).
\newblock


\bibitem[\protect\citeauthoryear{CodeCombat}{CodeCombat}{2014}]%
        {codecombat}
\bibfield{author}{\bibinfo{person}{Inc CodeCombat}.}
  \bibinfo{year}{2014}\natexlab{}.
\newblock \bibinfo{title}{CodeCombat}.
\newblock
\newblock
\urldef\tempurl%
\url{https://codecombat.com/}
\showURL{%
\tempurl}


\bibitem[\protect\citeauthoryear{CodeMonkey~Studios}{CodeMonkey~Studios}{2014}]%
        {codemonkey}
\bibfield{author}{\bibinfo{person}{Inc CodeMonkey~Studios}.}
  \bibinfo{year}{2014}\natexlab{}.
\newblock \bibinfo{title}{CodeMonkey, programming game}.
\newblock
\newblock
\urldef\tempurl%
\url{https://www.codemonkey.com/}
\showURL{%
\tempurl}


\bibitem[\protect\citeauthoryear{CodinGame}{CodinGame}{2015}]%
        {codingame}
\bibfield{author}{\bibinfo{person}{CodinGame}.}
  \bibinfo{year}{2015}\natexlab{}.
\newblock \bibinfo{title}{CodinGame}.
\newblock
\newblock
\urldef\tempurl%
\url{https://www.codingame.com/}
\showURL{%
\tempurl}


\bibitem[\protect\citeauthoryear{Corporation}{Corporation}{2015}]%
        {humanresource}
\bibfield{author}{\bibinfo{person}{Tomorrow Corporation}.}
  \bibinfo{year}{2015}\natexlab{}.
\newblock \bibinfo{title}{Human Resource Machine}.
\newblock
\newblock
\urldef\tempurl%
\url{https://tomorrowcorporation.com/humanresourcemachine}
\showURL{%
\tempurl}


\bibitem[\protect\citeauthoryear{Dieker, Hughes, Hynes, and Straub}{Dieker
  et~al\mbox{.}}{2017}]%
        {dieker2017using}
\bibfield{author}{\bibinfo{person}{Lisa~A Dieker}, \bibinfo{person}{Charles~E
  Hughes}, \bibinfo{person}{Michael~C Hynes}, {and} \bibinfo{person}{Carrie
  Straub}.} \bibinfo{year}{2017}\natexlab{}.
\newblock \showarticletitle{Using simulated virtual environments to improve
  teacher performance}.
\newblock \bibinfo{journal}{\emph{School University Partnerships (Journal of
  the National Association for Professional Development Schools): Special
  Issue: Technology to Enhance PDS}} \bibinfo{volume}{10}, \bibinfo{number}{3}
  (\bibinfo{year}{2017}), \bibinfo{pages}{62--81}.
\newblock


\bibitem[\protect\citeauthoryear{Edwards}{Edwards}{2004}]%
        {edwards2004using}
\bibfield{author}{\bibinfo{person}{Stephen~H Edwards}.}
  \bibinfo{year}{2004}\natexlab{}.
\newblock \showarticletitle{Using software testing to move students from
  trial-and-error to reflection-in-action}. In
  \bibinfo{booktitle}{\emph{Proceedings of the 35th SIGCSE technical symposium
  on Computer science education}}. \bibinfo{pages}{26--30}.
\newblock


\bibitem[\protect\citeauthoryear{Fekete, Kay, Kingston, and Wimalaratne}{Fekete
  et~al\mbox{.}}{2000}]%
        {fekete2000supporting}
\bibfield{author}{\bibinfo{person}{Alan Fekete}, \bibinfo{person}{Judy Kay},
  \bibinfo{person}{Jeff Kingston}, {and} \bibinfo{person}{Kapila Wimalaratne}.}
  \bibinfo{year}{2000}\natexlab{}.
\newblock \showarticletitle{Supporting reflection in introductory computer
  science}.
\newblock \bibinfo{journal}{\emph{ACM SIGCSE Bulletin}} \bibinfo{volume}{32},
  \bibinfo{number}{1} (\bibinfo{year}{2000}), \bibinfo{pages}{144--148}.
\newblock


\bibitem[\protect\citeauthoryear{Fleck and Fitzpatrick}{Fleck and
  Fitzpatrick}{2006}]%
        {fleck2006supporting}
\bibfield{author}{\bibinfo{person}{Rowanne Fleck} {and}
  \bibinfo{person}{Geraldine Fitzpatrick}.} \bibinfo{year}{2006}\natexlab{}.
\newblock \showarticletitle{Supporting collaborative reflection with passive
  image capture}.
\newblock  (\bibinfo{year}{2006}).
\newblock


\bibitem[\protect\citeauthoryear{Fleck and Fitzpatrick}{Fleck and
  Fitzpatrick}{2010}]%
        {fleck2010reflecting}
\bibfield{author}{\bibinfo{person}{Rowanne Fleck} {and}
  \bibinfo{person}{Geraldine Fitzpatrick}.} \bibinfo{year}{2010}\natexlab{}.
\newblock \showarticletitle{Reflecting on reflection: framing a design
  landscape}. In \bibinfo{booktitle}{\emph{Proceedings of the 22nd Conference
  of the Computer-Human Interaction Special Interest Group of Australia on
  Computer-Human Interaction}}. ACM, \bibinfo{pages}{216--223}.
\newblock


\bibitem[\protect\citeauthoryear{Games}{Games}{2010}]%
        {manufactoria}
\bibfield{author}{\bibinfo{person}{PleasingFungus Games}.}
  \bibinfo{year}{2010}\natexlab{}.
\newblock \bibinfo{title}{Manufactoria}.
\newblock
\newblock
\urldef\tempurl%
\url{http://pleasingfungus.com/Manufactoria/}
\showURL{%
\tempurl}


\bibitem[\protect\citeauthoryear{Garris, Ahlers, and Driskell}{Garris
  et~al\mbox{.}}{2002}]%
        {garris2002games}
\bibfield{author}{\bibinfo{person}{Rosemary Garris}, \bibinfo{person}{Robert
  Ahlers}, {and} \bibinfo{person}{James~E Driskell}.}
  \bibinfo{year}{2002}\natexlab{}.
\newblock \showarticletitle{Games, motivation, and learning: A research and
  practice model}.
\newblock \bibinfo{journal}{\emph{Simulation \& gaming}} \bibinfo{volume}{33},
  \bibinfo{number}{4} (\bibinfo{year}{2002}), \bibinfo{pages}{441--467}.
\newblock


\bibitem[\protect\citeauthoryear{Gee}{Gee}{2003}]%
        {gee2003video}
\bibfield{author}{\bibinfo{person}{James~Paul Gee}.}
  \bibinfo{year}{2003}\natexlab{}.
\newblock \showarticletitle{What video games have to teach us about learning
  and literacy}.
\newblock \bibinfo{journal}{\emph{Computers in Entertainment (CIE)}}
  \bibinfo{volume}{1}, \bibinfo{number}{1} (\bibinfo{year}{2003}),
  \bibinfo{pages}{20--20}.
\newblock


\bibitem[\protect\citeauthoryear{Giannakos}{Giannakos}{2013}]%
        {giannakos2013enjoy}
\bibfield{author}{\bibinfo{person}{Michail~N Giannakos}.}
  \bibinfo{year}{2013}\natexlab{}.
\newblock \showarticletitle{Enjoy and learn with educational games: Examining
  factors affecting learning performance}.
\newblock \bibinfo{journal}{\emph{Computers \& Education}}
  \bibinfo{volume}{68} (\bibinfo{year}{2013}), \bibinfo{pages}{429--439}.
\newblock


\bibitem[\protect\citeauthoryear{Honey and Hilton}{Honey and Hilton}{2011}]%
        {honey2011learning}
\bibfield{author}{\bibinfo{person}{Margaret~A Honey} {and}
  \bibinfo{person}{Margaret~L Hilton}.} \bibinfo{year}{2011}\natexlab{}.
\newblock \showarticletitle{Learning science through computer games}.
\newblock \bibinfo{journal}{\emph{National Academies Press, Washington, DC}}
  (\bibinfo{year}{2011}).
\newblock


\bibitem[\protect\citeauthoryear{Hooshyar, Yousefi, Wang, and Lim}{Hooshyar
  et~al\mbox{.}}{2018}]%
        {hooshyar2018data}
\bibfield{author}{\bibinfo{person}{Danial Hooshyar}, \bibinfo{person}{Moslem
  Yousefi}, \bibinfo{person}{Minhong Wang}, {and} \bibinfo{person}{Heuiseok
  Lim}.} \bibinfo{year}{2018}\natexlab{}.
\newblock \showarticletitle{A data-driven procedural-content-generation
  approach for educational games}.
\newblock \bibinfo{journal}{\emph{Journal of Computer Assisted Learning}}
  \bibinfo{volume}{34}, \bibinfo{number}{6} (\bibinfo{year}{2018}),
  \bibinfo{pages}{731--739}.
\newblock


\bibitem[\protect\citeauthoryear{Hughes, Hall, Ingraham, Epstein, and
  Hughes}{Hughes et~al\mbox{.}}{2016}]%
        {hughes2016enhancing}
\bibfield{author}{\bibinfo{person}{Charles~E Hughes}, \bibinfo{person}{Thomas
  Hall}, \bibinfo{person}{Kathleen Ingraham}, \bibinfo{person}{Jennifer~A
  Epstein}, {and} \bibinfo{person}{Darin~E Hughes}.}
  \bibinfo{year}{2016}\natexlab{}.
\newblock \showarticletitle{Enhancing protective role-playing behaviors through
  avatar-based scenarios}. In \bibinfo{booktitle}{\emph{2016 IEEE International
  Conference on Serious Games and Applications for Health (SeGAH)}}. IEEE,
  \bibinfo{pages}{1--6}.
\newblock


\bibitem[\protect\citeauthoryear{Left and Viana}{Left and Viana}{2012}]%
        {cargobot}
\bibfield{author}{\bibinfo{person}{Two~Lives Left} {and} \bibinfo{person}{Rui
  Viana}.} \bibinfo{year}{2012}\natexlab{}.
\newblock \bibinfo{title}{Cargo-Bot}.
\newblock
\newblock
\urldef\tempurl%
\url{https://twolivesleft.com/CargoBot/}
\showURL{%
\tempurl}


\bibitem[\protect\citeauthoryear{Light~Bot}{Light~Bot}{2008}]%
        {lightbot}
\bibfield{author}{\bibinfo{person}{Inc Light~Bot}.}
  \bibinfo{year}{2008}\natexlab{}.
\newblock \bibinfo{title}{Light Bot}.
\newblock
\newblock
\urldef\tempurl%
\url{https://lightbot.com/flash.html}
\showURL{%
\tempurl}


\bibitem[\protect\citeauthoryear{Lin, Hmelo, Kinzer, and Secules}{Lin
  et~al\mbox{.}}{1999}]%
        {lin1999designing}
\bibfield{author}{\bibinfo{person}{Xiaodong Lin}, \bibinfo{person}{Cindy
  Hmelo}, \bibinfo{person}{Charles~K Kinzer}, {and} \bibinfo{person}{Teresa~J
  Secules}.} \bibinfo{year}{1999}\natexlab{}.
\newblock \showarticletitle{Designing technology to support reflection}.
\newblock \bibinfo{journal}{\emph{Educational Technology Research and
  Development}} \bibinfo{volume}{47}, \bibinfo{number}{3}
  (\bibinfo{year}{1999}), \bibinfo{pages}{43--62}.
\newblock


\bibitem[\protect\citeauthoryear{Maciuszek and Martens}{Maciuszek and
  Martens}{2011}]%
        {maciuszek2011computer}
\bibfield{author}{\bibinfo{person}{Dennis Maciuszek} {and}
  \bibinfo{person}{Alke Martens}.} \bibinfo{year}{2011}\natexlab{}.
\newblock \showarticletitle{Computer role-playing games as an educational game
  genre: Activities and reflection}. In \bibinfo{booktitle}{\emph{European
  Conference on Games Based Learning}}. Academic Conferences International
  Limited, \bibinfo{pages}{368--500}.
\newblock


\bibitem[\protect\citeauthoryear{Maze}{Maze}{2018}]%
        {blockmaze}
\bibfield{author}{\bibinfo{person}{Blockly~Games Maze}.}
  \bibinfo{year}{2018}\natexlab{}.
\newblock \bibinfo{title}{Blockly: Maze, programming game}.
\newblock
\newblock
\urldef\tempurl%
\url{https://blockly.games/maze}
\showURL{%
\tempurl}


\bibitem[\protect\citeauthoryear{Moon}{Moon}{2013}]%
        {moon2013reflection}
\bibfield{author}{\bibinfo{person}{Jennifer~A Moon}.}
  \bibinfo{year}{2013}\natexlab{}.
\newblock \bibinfo{booktitle}{\emph{Reflection in learning and professional
  development: Theory and practice}}.
\newblock \bibinfo{publisher}{Routledge}.
\newblock


\bibitem[\protect\citeauthoryear{Moreno and Mayer}{Moreno and Mayer}{2005}]%
        {moreno2005role}
\bibfield{author}{\bibinfo{person}{Roxana Moreno} {and}
  \bibinfo{person}{Richard~E Mayer}.} \bibinfo{year}{2005}\natexlab{}.
\newblock \showarticletitle{Role of guidance, reflection, and interactivity in
  an agent-based multimedia game.}
\newblock \bibinfo{journal}{\emph{Journal of educational psychology}}
  \bibinfo{volume}{97}, \bibinfo{number}{1} (\bibinfo{year}{2005}),
  \bibinfo{pages}{117}.
\newblock


\bibitem[\protect\citeauthoryear{Mott, Callaway, Zettlemoyer, Lee, and
  Lester}{Mott et~al\mbox{.}}{1999}]%
        {mott1999towards}
\bibfield{author}{\bibinfo{person}{Bradford~W Mott}, \bibinfo{person}{Charles~B
  Callaway}, \bibinfo{person}{Luke~S Zettlemoyer}, \bibinfo{person}{Seung~Y
  Lee}, {and} \bibinfo{person}{James~C Lester}.}
  \bibinfo{year}{1999}\natexlab{}.
\newblock \showarticletitle{Towards narrative-centered learning environments}.
  In \bibinfo{booktitle}{\emph{Proceedings of the 1999 AAAI fall symposium on
  narrative intelligence}}. \bibinfo{pages}{78--82}.
\newblock


\bibitem[\protect\citeauthoryear{Multi-Dimensional}{Multi-Dimensional}{2015}]%
        {codespells}
\bibfield{author}{\bibinfo{person}{Multi-Dimensional}.}
  \bibinfo{year}{2015}\natexlab{}.
\newblock \bibinfo{title}{Code Spells}.
\newblock
\newblock
\urldef\tempurl%
\url{https://codespells.org/}
\showURL{%
\tempurl}


\bibitem[\protect\citeauthoryear{Ostrovsky}{Ostrovsky}{[n.d.]}]%
        {roboZZle}
\bibfield{author}{\bibinfo{person}{Igor Ostrovsky}.}
  \bibinfo{year}{[n.d.]}\natexlab{}.
\newblock \bibinfo{title}{RoboZZle}.
\newblock
\newblock
\urldef\tempurl%
\url{http://robozzle.com/}
\showURL{%
\tempurl}


\bibitem[\protect\citeauthoryear{O’Neil, Chung, Kerr, Vendlinski, Buschang,
  and Mayer}{O’Neil et~al\mbox{.}}{2014}]%
        {o2014adding}
\bibfield{author}{\bibinfo{person}{Harold~F O’Neil},
  \bibinfo{person}{Gregory~KWK Chung}, \bibinfo{person}{Deirdre Kerr},
  \bibinfo{person}{Terry~P Vendlinski}, \bibinfo{person}{Rebecca~E Buschang},
  {and} \bibinfo{person}{Richard~E Mayer}.} \bibinfo{year}{2014}\natexlab{}.
\newblock \showarticletitle{Adding self-explanation prompts to an educational
  computer game}.
\newblock \bibinfo{journal}{\emph{Computers in Human Behavior}}
  \bibinfo{volume}{30} (\bibinfo{year}{2014}), \bibinfo{pages}{23--28}.
\newblock


\bibitem[\protect\citeauthoryear{Peirce, Conlan, and Wade}{Peirce
  et~al\mbox{.}}{2008}]%
        {peirce2008adaptive}
\bibfield{author}{\bibinfo{person}{Neil Peirce}, \bibinfo{person}{Owen Conlan},
  {and} \bibinfo{person}{Vincent Wade}.} \bibinfo{year}{2008}\natexlab{}.
\newblock \showarticletitle{Adaptive educational games: Providing non-invasive
  personalised learning experiences}. In \bibinfo{booktitle}{\emph{2008 second
  IEEE international conference on digital game and intelligent toy enhanced
  learning}}. IEEE, \bibinfo{pages}{28--35}.
\newblock


\bibitem[\protect\citeauthoryear{Prensky}{Prensky}{2003}]%
        {prensky2003digital}
\bibfield{author}{\bibinfo{person}{Marc Prensky}.}
  \bibinfo{year}{2003}\natexlab{}.
\newblock \showarticletitle{Digital game-based learning}.
\newblock \bibinfo{journal}{\emph{Computers in Entertainment (CIE)}}
  \bibinfo{volume}{1}, \bibinfo{number}{1} (\bibinfo{year}{2003}),
  \bibinfo{pages}{21--21}.
\newblock


\bibitem[\protect\citeauthoryear{Randel, Morris, Wetzel, and Whitehill}{Randel
  et~al\mbox{.}}{1992}]%
        {randel1992effectiveness}
\bibfield{author}{\bibinfo{person}{Josephine~M Randel},
  \bibinfo{person}{Barbara~A Morris}, \bibinfo{person}{C~Douglas Wetzel}, {and}
  \bibinfo{person}{Betty~V Whitehill}.} \bibinfo{year}{1992}\natexlab{}.
\newblock \showarticletitle{The effectiveness of games for educational
  purposes: A review of recent research}.
\newblock \bibinfo{journal}{\emph{Simulation \& gaming}} \bibinfo{volume}{23},
  \bibinfo{number}{3} (\bibinfo{year}{1992}), \bibinfo{pages}{261--276}.
\newblock


\bibitem[\protect\citeauthoryear{Rowe, Mott, McQuiggan, Robison, Lee, and
  Lester}{Rowe et~al\mbox{.}}{2009}]%
        {rowe2009crystal}
\bibfield{author}{\bibinfo{person}{Jonathan Rowe}, \bibinfo{person}{Bradford
  Mott}, \bibinfo{person}{Scott McQuiggan}, \bibinfo{person}{Jennifer Robison},
  \bibinfo{person}{Sunyoung Lee}, {and} \bibinfo{person}{James Lester}.}
  \bibinfo{year}{2009}\natexlab{}.
\newblock \showarticletitle{Crystal island: A narrative-centered learning
  environment for eighth grade microbiology}. In
  \bibinfo{booktitle}{\emph{workshop on intelligent educational games at the
  14th international conference on artificial intelligence in education,
  Brighton, UK}}. \bibinfo{pages}{11--20}.
\newblock


\bibitem[\protect\citeauthoryear{Sabourin and Lester}{Sabourin and
  Lester}{2013}]%
        {sabourin2013affect}
\bibfield{author}{\bibinfo{person}{Jennifer~L Sabourin} {and}
  \bibinfo{person}{James~C Lester}.} \bibinfo{year}{2013}\natexlab{}.
\newblock \showarticletitle{Affect and engagement in Game-BasedLearning
  environments}.
\newblock \bibinfo{journal}{\emph{IEEE Transactions on Affective Computing}}
  \bibinfo{volume}{5}, \bibinfo{number}{1} (\bibinfo{year}{2013}),
  \bibinfo{pages}{45--56}.
\newblock


\bibitem[\protect\citeauthoryear{Sch{\"{o}}n}{Sch{\"{o}}n}{1983}]%
        {Schon1983}
\bibfield{author}{\bibinfo{person}{Donald~A Sch{\"{o}}n}.}
  \bibinfo{year}{1983}\natexlab{}.
\newblock \bibinfo{title}{{The reflective practitioner: how professionals think
  in action}}.
\newblock
\newblock
\showISBNx{046506874X;9780465068746;}


\bibitem[\protect\citeauthoryear{Schunk}{Schunk}{2012}]%
        {schunk2012learning}
\bibfield{author}{\bibinfo{person}{Dale~H Schunk}.}
  \bibinfo{year}{2012}\natexlab{}.
\newblock \bibinfo{booktitle}{\emph{Learning theories an educational
  perspective sixth edition}}.
\newblock \bibinfo{publisher}{Pearson}.
\newblock


\bibitem[\protect\citeauthoryear{Sengers, Boehner, David, and Kaye}{Sengers
  et~al\mbox{.}}{2005}]%
        {sengers2005reflective}
\bibfield{author}{\bibinfo{person}{Phoebe Sengers}, \bibinfo{person}{Kirsten
  Boehner}, \bibinfo{person}{Shay David}, {and} \bibinfo{person}{Joseph'Jofish'
  Kaye}.} \bibinfo{year}{2005}\natexlab{}.
\newblock \showarticletitle{Reflective design}. In
  \bibinfo{booktitle}{\emph{Proceedings of the 4th decennial conference on
  Critical computing: between sense and sensibility}}. ACM,
  \bibinfo{pages}{49--58}.
\newblock


\bibitem[\protect\citeauthoryear{Spires, Rowe, Mott, and Lester}{Spires
  et~al\mbox{.}}{2011}]%
        {spires2011problem}
\bibfield{author}{\bibinfo{person}{Hiller~A Spires},
  \bibinfo{person}{Jonathan~P Rowe}, \bibinfo{person}{Bradford~W Mott}, {and}
  \bibinfo{person}{James~C Lester}.} \bibinfo{year}{2011}\natexlab{}.
\newblock \showarticletitle{Problem solving and game-based learning: Effects of
  middle grade students' hypothesis testing strategies on learning outcomes}.
\newblock \bibinfo{journal}{\emph{Journal of Educational Computing Research}}
  \bibinfo{volume}{44}, \bibinfo{number}{4} (\bibinfo{year}{2011}),
  \bibinfo{pages}{453--472}.
\newblock


\bibitem[\protect\citeauthoryear{Valls-Vargas, Ontan{\'o}n, and
  Zhu}{Valls-Vargas et~al\mbox{.}}{2015}]%
        {valls2015exploring}
\bibfield{author}{\bibinfo{person}{Josep Valls-Vargas},
  \bibinfo{person}{Santiago Ontan{\'o}n}, {and} \bibinfo{person}{Jichen Zhu}.}
  \bibinfo{year}{2015}\natexlab{}.
\newblock \showarticletitle{Exploring player trace segmentation for dynamic
  play style prediction}. In \bibinfo{booktitle}{\emph{Eleventh Artificial
  Intelligence and Interactive Digital Entertainment Conference}}.
\newblock


\bibitem[\protect\citeauthoryear{Valls-Vargas, Zhu, and
  Onta{\~n}{\'o}n}{Valls-Vargas et~al\mbox{.}}{2017}]%
        {valls2017graph}
\bibfield{author}{\bibinfo{person}{Josep Valls-Vargas}, \bibinfo{person}{Jichen
  Zhu}, {and} \bibinfo{person}{Santiago Onta{\~n}{\'o}n}.}
  \bibinfo{year}{2017}\natexlab{}.
\newblock \showarticletitle{Graph grammar-based controllable generation of
  puzzles for a learning game about parallel programming}. In
  \bibinfo{booktitle}{\emph{Proceedings of the 12th International Conference on
  the Foundations of Digital Games}}. \bibinfo{pages}{1--10}.
\newblock


\bibitem[\protect\citeauthoryear{Wouters, Van~Nimwegen, Van~Oostendorp, and Van
  Der~Spek}{Wouters et~al\mbox{.}}{2013}]%
        {wouters2013meta}
\bibfield{author}{\bibinfo{person}{Pieter Wouters}, \bibinfo{person}{Christof
  Van~Nimwegen}, \bibinfo{person}{Herre Van~Oostendorp}, {and}
  \bibinfo{person}{Erik~D Van Der~Spek}.} \bibinfo{year}{2013}\natexlab{}.
\newblock \showarticletitle{A meta-analysis of the cognitive and motivational
  effects of serious games.}
\newblock \bibinfo{journal}{\emph{Journal of educational psychology}}
  \bibinfo{volume}{105}, \bibinfo{number}{2} (\bibinfo{year}{2013}),
  \bibinfo{pages}{249}.
\newblock


\bibitem[\protect\citeauthoryear{Yusoff, Crowder, Gilbert, and Wills}{Yusoff
  et~al\mbox{.}}{2009}]%
        {yusoff2009conceptual}
\bibfield{author}{\bibinfo{person}{Amri Yusoff}, \bibinfo{person}{Richard
  Crowder}, \bibinfo{person}{Lester Gilbert}, {and} \bibinfo{person}{Gary
  Wills}.} \bibinfo{year}{2009}\natexlab{}.
\newblock \showarticletitle{A conceptual framework for serious games}. In
  \bibinfo{booktitle}{\emph{2009 Ninth IEEE International Conference on
  Advanced Learning Technologies}}. IEEE, \bibinfo{pages}{21--23}.
\newblock


\bibitem[\protect\citeauthoryear{Zachtronics}{Zachtronics}{2011}]%
        {spacechem}
\bibfield{author}{\bibinfo{person}{Zachtronics}.}
  \bibinfo{year}{2011}\natexlab{}.
\newblock \bibinfo{title}{SpaceChem}.
\newblock
\newblock
\urldef\tempurl%
\url{http://www.zachtronics.com/spacechem/}
\showURL{%
\tempurl}


\bibitem[\protect\citeauthoryear{Zhu, Alderfer, Furqan, Nebolsky, Char, Smith,
  Villareale, and Onta{\~n}{\'o}n}{Zhu et~al\mbox{.}}{2019}]%
        {zhu2019programming}
\bibfield{author}{\bibinfo{person}{Jichen Zhu}, \bibinfo{person}{Katelyn
  Alderfer}, \bibinfo{person}{Anushay Furqan}, \bibinfo{person}{Jessica
  Nebolsky}, \bibinfo{person}{Bruce Char}, \bibinfo{person}{Brian Smith},
  \bibinfo{person}{Jennifer Villareale}, {and} \bibinfo{person}{Santiago
  Onta{\~n}{\'o}n}.} \bibinfo{year}{2019}\natexlab{}.
\newblock \showarticletitle{Programming in game space: how to represent
  parallel programming concepts in an educational game}. In
  \bibinfo{booktitle}{\emph{Proceedings of the 14th International Conference on
  the Foundations of Digital Games}}. \bibinfo{pages}{1--10}.
\newblock


\end{thebibliography}




 {\em Blockly: Maze}, {\em Cargo-Bot}, {\em Code Combat}, {\em CodeMonkey}, {\em Code Spells}, {\em CodinGame}, {\em Human Resource Machine}, {\em Light Bolt}, {\em Manufactoria}, {\em Parallel}, {\em RoboZZle} and {\em SpaceChem}
 

 
   @misc{blockmaze,
  title={Blockly: Maze, programming game},
  author={Blockly Games Maze},
  year={2018},
  url={https://blockly.games/maze}
}

   @misc{cargobot,
  title={Cargo-Bot},
  author={Two Lives Left and Rui Viana},
  year={2012},
  url={https://twolivesleft.com/CargoBot/}
}

   @misc{codecombat,
  title={CodeCombat},
  author={CodeCombat, Inc},
  year={2014},
  url={https://codecombat.com/}
}

   @misc{codemonkey,
  title={CodeMonkey, programming game},
  author={CodeMonkey Studios, Inc},
  year={2014},
  url={https://www.codemonkey.com/}
}

   @misc{codespells,
  title={Code Spells},
  author={Multi-Dimensional},
  year={2015},
  url={https://codespells.org/}
}

   @misc{codingame,
  title={CodinGame},
  author={CodinGame},
  year={2015},
  url={https://www.codingame.com/}
}

   @misc{humanresource,
  title={Human Resource Machine},
  author={Tomorrow Corporation},
  year={2015},
  url={https://tomorrowcorporation.com/humanresourcemachine}
}

   @misc{lightbot,
  title={Light Bot},
  author={Light Bot, Inc},
  year={2008},
  url={https://lightbot.com/flash.html}
}

   @misc{manufactoria,
  title={Manufactoria},
  author={PleasingFungus Games},
  year={2010},
  url={http://pleasingfungus.com/Manufactoria/}
}

   @misc{roboZZle,
  title={RoboZZle},
  author={Igor Ostrovsky},
  url={http://robozzle.com/}
}

   @misc{spacechem,
  title={SpaceChem},
  author={Zachtronics},
  year={2011},
  url={http://www.zachtronics.com/spacechem/}
}



@misc{Schon1983,
address = {New York},
annote = {HD8038.A1 S35 1983},
author = {Sch{\"{o}}n, Donald A},
isbn = {046506874X;9780465068746;},
keywords = {Career patterns,Careers,Introspection (Theory of knowledge),Jobs,Knowledge,Knowledge of self,Mind,Professional services,Professions,Reflection (Theory of knowledge),Reflexive,Reflexive knowledge,Self-knowledge,Theory of,Thinking,Thought and thinking,Thoughts},
publisher = {Basic Books},
title = {{The reflective practitioner: how professionals think in action}},
year = {1983}
}

@inproceedings{mott1999towards,
  title={Towards narrative-centered learning environments},
  author={Mott, Bradford W and Callaway, Charles B and Zettlemoyer, Luke S and Lee, Seung Y and Lester, James C},
  booktitle={Proceedings of the 1999 AAAI fall symposium on narrative intelligence},
  pages={78--82},
  year={1999}
}



@article{sabourin2013affect,
  title={Affect and engagement in Game-BasedLearning environments},
  author={Sabourin, Jennifer L and Lester, James C},
  journal={IEEE Transactions on Affective Computing},
  volume={5},
  number={1},
  pages={45--56},
  year={2013},
  publisher={IEEE}
}



@inproceedings{ha2011goal,
  title={Goal recognition with Markov logic networks for player-adaptive games},
  author={Ha, Eun Young and Rowe, Jonathan P and Mott, Bradford W and Lester, James C},
  booktitle={Seventh Artificial Intelligence and Interactive Digital Entertainment Conference},
  year={2011}
}

@inproceedings{rowe2009crystal,
  title={Crystal island: A narrative-centered learning environment for eighth grade microbiology},
  author={Rowe, Jonathan and Mott, Bradford and McQuiggan, Scott and Robison, Jennifer and Lee, Sunyoung and Lester, James},
  booktitle={workshop on intelligent educational games at the 14th international conference on artificial intelligence in education, Brighton, UK},
  pages={11--20},
  year={2009}
}


@inproceedings{peirce2008adaptive,
  title={Adaptive educational games: Providing non-invasive personalised learning experiences},
  author={Peirce, Neil and Conlan, Owen and Wade, Vincent},
  booktitle={2008 second IEEE international conference on digital game and intelligent toy enhanced learning},
  pages={28--35},
  year={2008},
  organization={IEEE}
}


@article{spires2011problem,
  title={Problem solving and game-based learning: Effects of middle grade students' hypothesis testing strategies on learning outcomes},
  author={Spires, Hiller A and Rowe, Jonathan P and Mott, Bradford W and Lester, James C},
  journal={Journal of Educational Computing Research},
  volume={44},
  number={4},
  pages={453--472},
  year={2011},
  publisher={SAGE Publications Sage CA: Los Angeles, CA}
}



@article{wouters2013meta,
  title={A meta-analysis of the cognitive and motivational effects of serious games.},
  author={Wouters, Pieter and Van Nimwegen, Christof and Van Oostendorp, Herre and Van Der Spek, Erik D},
  journal={Journal of educational psychology},
  volume={105},
  number={2},
  pages={249},
  year={2013},
  publisher={American Psychological Association}
}


@article{giannakos2013enjoy,
  title={Enjoy and learn with educational games: Examining factors affecting learning performance},
  author={Giannakos, Michail N},
  journal={Computers \& Education},
  volume={68},
  pages={429--439},
  year={2013},
  publisher={Elsevier}
}


@article{hooshyar2018data,
  title={A data-driven procedural-content-generation approach for educational games},
  author={Hooshyar, Danial and Yousefi, Moslem and Wang, Minhong and Lim, Heuiseok},
  journal={Journal of Computer Assisted Learning},
  volume={34},
  number={6},
  pages={731--739},
  year={2018},
  publisher={Wiley Online Library}
}


@inproceedings{valls2017graph,
  title={Graph grammar-based controllable generation of puzzles for a learning game about parallel programming},
  author={Valls-Vargas, Josep and Zhu, Jichen and Onta{\~n}{\'o}n, Santiago},
  booktitle={Proceedings of the 12th International Conference on the Foundations of Digital Games},
  pages={1--10},
  year={2017}
}


@article{gee2003video,
  title={What video games have to teach us about learning and literacy},
  author={Gee, James Paul},
  journal={Computers in Entertainment (CIE)},
  volume={1},
  number={1},
  pages={20--20},
  year={2003},
  publisher={ACM}
}

@article{moreno2005role,
  title={Role of guidance, reflection, and interactivity in an agent-based multimedia game.},
  author={Moreno, Roxana and Mayer, Richard E},
  journal={Journal of educational psychology},
  volume={97},
  number={1},
  pages={117},
  year={2005},
  publisher={American Psychological Association}
}

@article{paras2005game,
  title={Game, motivation, and effective learning: An integrated model for educational game design},
  author={Paras, Brad},
  year={2005}
}

@inproceedings{alderfer2018lessons,
  title={Lessons Learned From an Interactive Educational Computer Game About Concurrent Programming},
  author={Alderfer, Katelyn Bright and Smith, Brian K and Onta{\~n}{\'o}n, Santiago and Char, Bruce and Nebolsky, Jessica and Zhu, Jichen and Furqan, Anushay and Freed, Evan and Patterson, Justin and Valls-Vargas, Josep},
  booktitle={Proceedings of the 49th ACM Technical Symposium on Computer Science Education},
  pages={1077--1077},
  year={2018}
}

@inproceedings{maciuszek2011computer,
  title={Computer role-playing games as an educational game genre: Activities and reflection},
  author={Maciuszek, Dennis and Martens, Alke},
  booktitle={European Conference on Games Based Learning},
  pages={368--500},
  year={2011},
  organization={Academic Conferences International Limited}
}

@inproceedings{yusoff2009conceptual,
  title={A conceptual framework for serious games},
  author={Yusoff, Amri and Crowder, Richard and Gilbert, Lester and Wills, Gary},
  booktitle={2009 Ninth IEEE International Conference on Advanced Learning Technologies},
  pages={21--23},
  year={2009},
  organization={IEEE}
}
@article{maragos2005towards,
  title={Towards the design of intelligent educational gaming systems},
  author={Maragos, Konstantinos and Grigoriadou, Maria},
  journal={Proc. AIED05 WORKSHOP5: Educational Games as Intelligent Learning Environments},
  pages={35--38},
  year={2005}
}
@article{garris2002games,
  title={Games, motivation, and learning: A research and practice model},
  author={Garris, Rosemary and Ahlers, Robert and Driskell, James E},
  journal={Simulation \& gaming},
  volume={33},
  number={4},
  pages={441--467},
  year={2002},
  publisher={Sage Publications Sage CA: Thousand Oaks, CA}
}
@article{emihovich1988effects,
  title={Effects of Logo and CAI on black first graders' achievement, reflectivity, and self-esteem},
  author={Emihovich, Catherine and Miller, Gloria E},
  journal={The Elementary School Journal},
  volume={88},
  number={5},
  pages={473--487},
  year={1988},
  publisher={University of Chicago Press}
}

@article{honey2011learning,
  title={Learning science through computer games},
  author={Honey, Margaret A and Hilton, Margaret L},
  journal={National Academies Press, Washington, DC},
  year={2011}
}

@book{schunk2012learning,
  title={Learning theories an educational perspective sixth edition},
  author={Schunk, Dale H},
  year={2012},
  publisher={Pearson}
}

@article{o2014adding,
  title={Adding self-explanation prompts to an educational computer game},
  author={O’Neil, Harold F and Chung, Gregory KWK and Kerr, Deirdre and Vendlinski, Terry P and Buschang, Rebecca E and Mayer, Richard E},
  journal={Computers in Human Behavior},
  volume={30},
  pages={23--28},
  year={2014},
  publisher={Elsevier}
}

@incollection{vanlehn1988toward,
  title={Toward a theory of impasse-driven learning},
  author={VanLehn, Kurt},
  booktitle={Learning issues for intelligent tutoring systems},
  pages={19--41},
  year={1988},
  publisher={Springer}
}

@article{prensky2003digital,
  title={Digital game-based learning},
  author={Prensky, Marc},
  journal={Computers in Entertainment (CIE)},
  volume={1},
  number={1},
  pages={21--21},
  year={2003},
  publisher={ACM}
}

@inproceedings{valls2015exploring,
  title={Exploring player trace segmentation for dynamic play style prediction},
  author={Valls-Vargas, Josep and Ontan{\'o}n, Santiago and Zhu, Jichen},
  booktitle={Eleventh Artificial Intelligence and Interactive Digital Entertainment Conference},
  year={2015}
}

@inproceedings{barnes2008game2learn,
  title={Game2Learn: improving the motivation of CS1 students},
  author={Barnes, Tiffany and Powell, Eve and Chaffin, Amanda and Lipford, Heather},
  booktitle={Proceedings of the 3rd international conference on Game development in computer science education},
  pages={1--5},
  year={2008},
  organization={ACM}
}

@inproceedings{cliburn2006effectiveness,
  title={The effectiveness of games as assignments in an introductory programming course},
  author={Cliburn, Daniel C},
  booktitle={Proceedings. Frontiers in Education. 36th Annual Conference},
  pages={6--10},
  year={2006},
  organization={IEEE}
}

@article{randel1992effectiveness,
  title={The effectiveness of games for educational purposes: A review of recent research},
  author={Randel, Josephine M and Morris, Barbara A and Wetzel, C Douglas and Whitehill, Betty V},
  journal={Simulation \& gaming},
  volume={23},
  number={3},
  pages={261--276},
  year={1992},
  publisher={Sage Publications Sage CA: Thousand Oaks, CA}
}

@article{lin1999designing,
  title={Designing technology to support reflection},
  author={Lin, Xiaodong and Hmelo, Cindy and Kinzer, Charles K and Secules, Teresa J},
  journal={Educational Technology Research and Development},
  volume={47},
  number={3},
  pages={43--62},
  year={1999},
  publisher={Springer}
}

@inproceedings{fleck2010reflecting,
  title={Reflecting on reflection: framing a design landscape},
  author={Fleck, Rowanne and Fitzpatrick, Geraldine},
  booktitle={Proceedings of the 22nd Conference of the Computer-Human Interaction Special Interest Group of Australia on Computer-Human Interaction},
  pages={216--223},
  year={2010},
  organization={ACM}
}

@inproceedings{edwards2004using,
  title={Using software testing to move students from trial-and-error to reflection-in-action},
  author={Edwards, Stephen H},
  booktitle={Proceedings of the 35th SIGCSE technical symposium on Computer science education},
  pages={26--30},
  year={2004}
}

@article{fekete2000supporting,
  title={Supporting reflection in introductory computer science},
  author={Fekete, Alan and Kay, Judy and Kingston, Jeff and Wimalaratne, Kapila},
  journal={ACM SIGCSE Bulletin},
  volume={32},
  number={1},
  pages={144--148},
  year={2000},
  publisher={ACM New York, NY, USA}
}

@article{fleck2006supporting,
  title={Supporting collaborative reflection with passive image capture},
  author={Fleck, Rowanne and Fitzpatrick, Geraldine},
  year={2006}
}

@inproceedings{sengers2005reflective,
  title={Reflective design},
  author={Sengers, Phoebe and Boehner, Kirsten and David, Shay and Kaye, Joseph'Jofish'},
  booktitle={Proceedings of the 4th decennial conference on Critical computing: between sense and sensibility},
  pages={49--58},
  year={2005},
  organization={ACM}
}

@article{boud1996promoting,
  title={Promoting reflection in learning: A model},
  author={Boud, David and Keogh, Rosemary and Walker, David},
  journal={Boundaries of adult learning},
  volume={1},
  pages={32--56},
  year={1996},
  publisher={Routledge, New York, NY}
}

@article{anderson2018failing,
  title={Failing up: How failure in a game environment promotes learning through discourse},
  author={Anderson, Craig G and Dalsen, Jen and Kumar, Vishesh and Berland, Matthew and Steinkuehler, Constance},
  journal={Thinking Skills and Creativity},
  volume={30},
  pages={135--144},
  year={2018},
  publisher={Elsevier}
}

@book{juul2013art,
  title={The art of failure: An essay on the pain of playing video games},
  author={Juul, Jesper},
  year={2013},
  publisher={MIT press}
}

@book{moon2013reflection,
  title={Reflection in learning and professional development: Theory and practice},
  author={Moon, Jennifer A},
  year={2013},
  publisher={Routledge}
}

@article{bransford1993ideal,
  title={The IDEAL problem solver},
  author={Bransford, John D and Stein, Barry S},
  year={1993}
}

@inproceedings{zhu2019programming,
  title={Programming in game space: how to represent parallel programming concepts in an educational game},
  author={Zhu, Jichen and Alderfer, Katelyn and Furqan, Anushay and Nebolsky, Jessica and Char, Bruce and Smith, Brian and Villareale, Jennifer and Onta{\~n}{\'o}n, Santiago},
  booktitle={Proceedings of the 14th International Conference on the Foundations of Digital Games},
  pages={1--10},
  year={2019}
}

@article{schon1938reflective,
  title={The reflective practitioner},
  author={Sch{\"o}n, Donald},
  journal={New York},
  volume={1083},
  year={1938}
}

@article{bizzocchi2011well,
  title={Well read: Applying close reading techniques to gameplay experiences},
  author={Bizzocchi, Jim and Tanenbaum, Joshua},
  journal={Well Played 3.0: Video Games, Value and Meaning},
  pages={262--290},
  year={2011},
  publisher={ETC Press Pittsburgh}
}


@article{dieker2017using,
  title={Using simulated virtual environments to improve teacher performance},
  author={Dieker, Lisa A and Hughes, Charles E and Hynes, Michael C and Straub, Carrie},
  journal={School University Partnerships (Journal of the National Association for Professional Development Schools): Special Issue: Technology to Enhance PDS},
  volume={10},
  number={3},
  pages={62--81},
  year={2017}
}

@article{barmaki2018embodiment,
  title={Embodiment analytics of practicing teachers in a virtual immersive environment},
  author={Barmaki, Roghayeh and Hughes, Charles E},
  journal={Journal of Computer Assisted Learning},
  volume={34},
  number={4},
  pages={387--396},
  year={2018},
  publisher={Wiley Online Library}
}

@inproceedings{hughes2016enhancing,
  title={Enhancing protective role-playing behaviors through avatar-based scenarios},
  author={Hughes, Charles E and Hall, Thomas and Ingraham, Kathleen and Epstein, Jennifer A and Hughes, Darin E},
  booktitle={2016 IEEE International Conference on Serious Games and Applications for Health (SeGAH)},
  pages={1--6},
  year={2016},
  organization={IEEE}
}

@inproceedings{barmaki2018gesturing,
  title={Gesturing and Embodiment in Teaching: Investigating the Nonverbal‎ Behavior of Teachers in a Virtual Rehearsal Environment‎},
  author={Barmaki, Roghayeh and Hughes, Charles},
  booktitle={Thirty-Second AAAI Conference on Artificial Intelligence},
  year={2018}
}

@article{darling2003keeping,
  title={Keeping good teachers: Why it matters, what leaders can do},
  author={Darling-Hammond, Linda},
  journal={Educational leadership},
  volume={60},
  number={8},
  pages={6--13},
  year={2003},
  publisher={ASCD ASSOCIATION FOR SUPERVISION AND}
}

@article{emihovich1988effects,
  title={Effects of Logo and CAI on black first graders' achievement, reflectivity, and self-esteem},
  author={Emihovich, Catherine and Miller, Gloria E},
  journal={The Elementary School Journal},
  volume={88},
  number={5},
  pages={473--487},
  year={1988},
  publisher={University of Chicago Press}
}
\end{document}